\documentclass[pra,superscriptaddress,onecolumn,nofootinbib]{revtex4}

\usepackage{graphicx,epsfig}
\usepackage{color}
\usepackage[usenames,dvipsnames]{xcolor}
\usepackage{amsmath,bbm,amssymb, amsthm}
\usepackage{dsfont} 
\usepackage{stmaryrd}
\def\bb{\begin{eqnarray}}
\def\ee{\end{eqnarray}}
\newcommand{\ket}[1]{| #1 \rangle}
\newcommand{\bra}[1]{\langle #1 |}

\newcommand{\moy}[1]{\left\langle #1 \right\rangle} 
\newcommand{\idop}{\mathds{1}}

\begin{document}

\title{An Interaction-Free Quantum Measurement-Driven Engine}


\author{Cyril Elouard}

\affiliation{Department of Physics and Astronomy, University of Rochester, Rochester, NY 14627, USA}

\email{cyril.elouard@gmail.com}

\author{Mordecai Waegell}

\affiliation{Institute for Quantum studies, Chapman University, Orange, CA 92866, USA}

\author{Benjamin Huard}

\affiliation{Laboratoire de Physique,  \'Ecole  Normale  Sup\'erieure  de  Lyon,  46  all\'ee  d’Italie,  69364 Lyon Cedex 7, France}

\author{Andrew N. Jordan}

\affiliation{Department of Physics and Astronomy, University of Rochester, Rochester, NY 14627, USA}

\begin{abstract}
Recently highly-efficient quantum engines were devised by exploiting the stochastic energy changes induced by quantum measurement. Here we show that such an engine can be based on an interaction-free measurement, in which the meter seemingly does not interact with the measured object. We use a modified version of the Elitzur-Vaidman bomb tester, an interferometric setup able to detect the presence of a bomb triggered by a single photon without exploding it. In our case, a quantum bomb subject to a gravitational force is initially in a superposition of being inside and outside one of the interferometer arms. 
We show that the bomb can be lifted without blowing up. This occurs when a photon traversing the interferometer is detected at a port that is always dark when the bomb is located outside the arm. The required potential energy is provided by the photon (which plays the role of the meter) even though it was not absorbed by the bomb. A natural interpretation is that the photon traveled through the arm which does not contain the bomb -- otherwise the bomb would have exploded -- but it implies the surprising conclusion that the energy exchange occurred at a distance despite a local interaction Hamiltonian. We use the weak value formalism to support this interpretation and find evidence of contextuality. Regardless of interpretation, this interaction-free quantum measurement engine is able to lift the most sensitive bomb without setting it off.
\end{abstract}

\maketitle

\section{Introduction}
\label{intro}

Interaction-free measurements are often presented as a way of gaining information.  Given prior knowledge, we can classically learn something nonlocally about a system by measuring where it is not.  This is simply an updating of our knowledge given prior ignorance. A quantum mechanical version was introduced by Elitzur and Vaidman \cite{Elitzur93}: They envisioned an ultra sensitive bomb that would explode whenever it encountered a photon. When this bomb was placed into one arm of a tuned interferometer, detecting a photon at the dark port, i.e. the port that would normally never fire, brought simultaneously two pieces of information. First, that the interference pattern was modified, revealing that one of the arms of the interferometer must have been blocked, therefore indicating the presence of the bomb, while not exploding it. Second, that the photon was not absorbed by the bomb, which seems only compatible with the photon traveling through the arm not containing the bomb, and consequently never interacting with the bomb.
This type of quantum interaction-free measurement has since been generalized to a wide variety of situations, such as counter-factual quantum computation \cite{Mitchison2001May} or communication \cite{Salih13}, seemingly paradoxical experiments \cite{Hardy92}, etc. Some of them were successfully implemented experimentally \cite{Kwiat1995Jun,Hosten2006Feb,Lundeen09,Yokota09,Liu2012Jul,Kong2015Aug}. In all such cases, information about objects or actions is obtained, seemingly without having interacted with them.

\begin{figure}[h!]
\begin{center}
\includegraphics[width=0.6\textwidth]{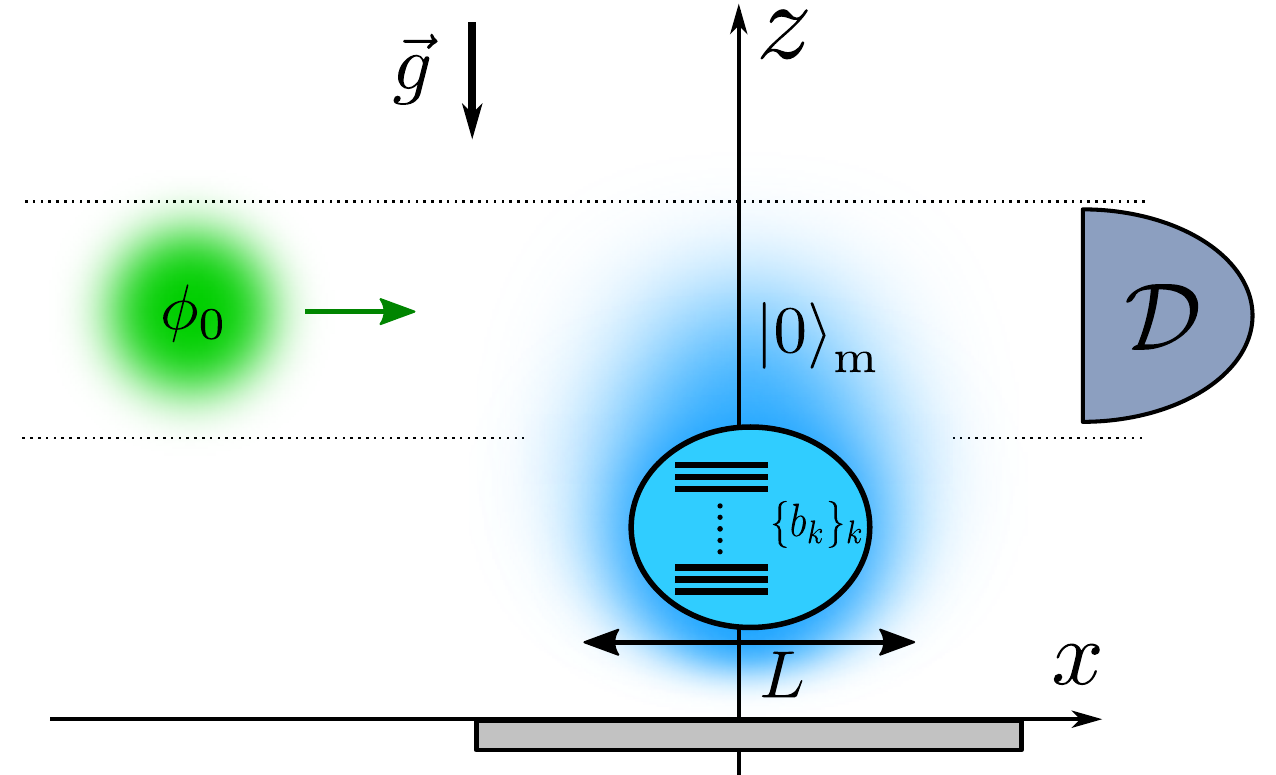}
\end{center}
\caption{Position measurement of a bomb using the transmission of a single photon. The bomb's motional degree of freedom is initially in the ground state $\ket{0}_\mathrm{m}$ of the quantum bouncing ball Hamiltonian, i.e. in the linear potential associated with the gravity field $\vec{g}$, and above a rigid platform enforcing its altitude to be $z>0$. The photon solely interacts with the part of the bomb's wavefuncion that overlaps with its own spatial wavefunction $\phi_0$ (see Appendix). An absorption of the photon is associated with a localization of the bomb's wavefunction within this overlap area.\label{f:Meas}}
\end{figure}

The purpose of the present paper is to show that one can change the energy of the measured objects -- playing the role of the bomb -- while performing an interaction-free measurement of their presence. To be specific, we propose a design of an engine able to do useful work on an Elitzur-Vaidman bomb, despite never having (seemingly) interacted with it.
In the setup we propose, the bomb has both an internal degree of freedom able to absorb the photon, and a motional degree of freedom, treated quantum mechanically, that is initially in a superposition of being inside and outside the probed area (see Fig.~\ref{f:Meas}). A click at the optical dark port ascertains the presence of the bomb as in the original Elitzur-Vaidman bomb, but in addition projects the bomb motional state inside the arm of the interferometer, which results in an increase of its potential energy. This energy variation can be cyclically extracted in an engine cycle powered by the measurement process, as proposed in \cite{Elouard17Maxwell,Elouard18Engine,Bresque20}.

To prove that such an engine can indeed be built, and unveil the mechanisms at the origin of this energy increase, we study a specific microscopic model of the bomb and the single photon interferometer, characterized by a local interaction Hamiltonian, and show that energy conservation holds during the measurement process, such that the energy gained by the bomb is provided by the photon. As in the original Elitzur-Vaidman setup, the only interpretation which seems to be compatible with both the photon being transmitted and the bomb being present in the interferometer is the photon traveling via the other arm. However, in our setup, this implies the surprising conclusion that the energy transfer between the photon and the bomb occurred at a distance, despite a local interaction Hamiltonian. 

In the first section, we present the model for the photon-bomb interaction and analyze the conditions leading to the bomb's position measurement. We then detail the energy balance. We show that the useful extracted work is associated with a red-shift of the photon frequency. 
In the second section, we consider that the photon's beam path is one of the two arms of an interferometer, and show that this allows us to implement a seemingly interaction-free position measurement on the atom. In the third section, we calculate the post-selected anomalous energy gains (weak-values), that we show support this interpretation, and bring some insights about how the energy exchange takes place. They also allow us to formally point out the contextual aspect of the phenomenon from the presence of anomalous weak-values. We finally discuss the role of energy uncertainty and coherence of the bomb.

\begin{figure}[h!]
\begin{center}
\includegraphics[width=0.7\textwidth]{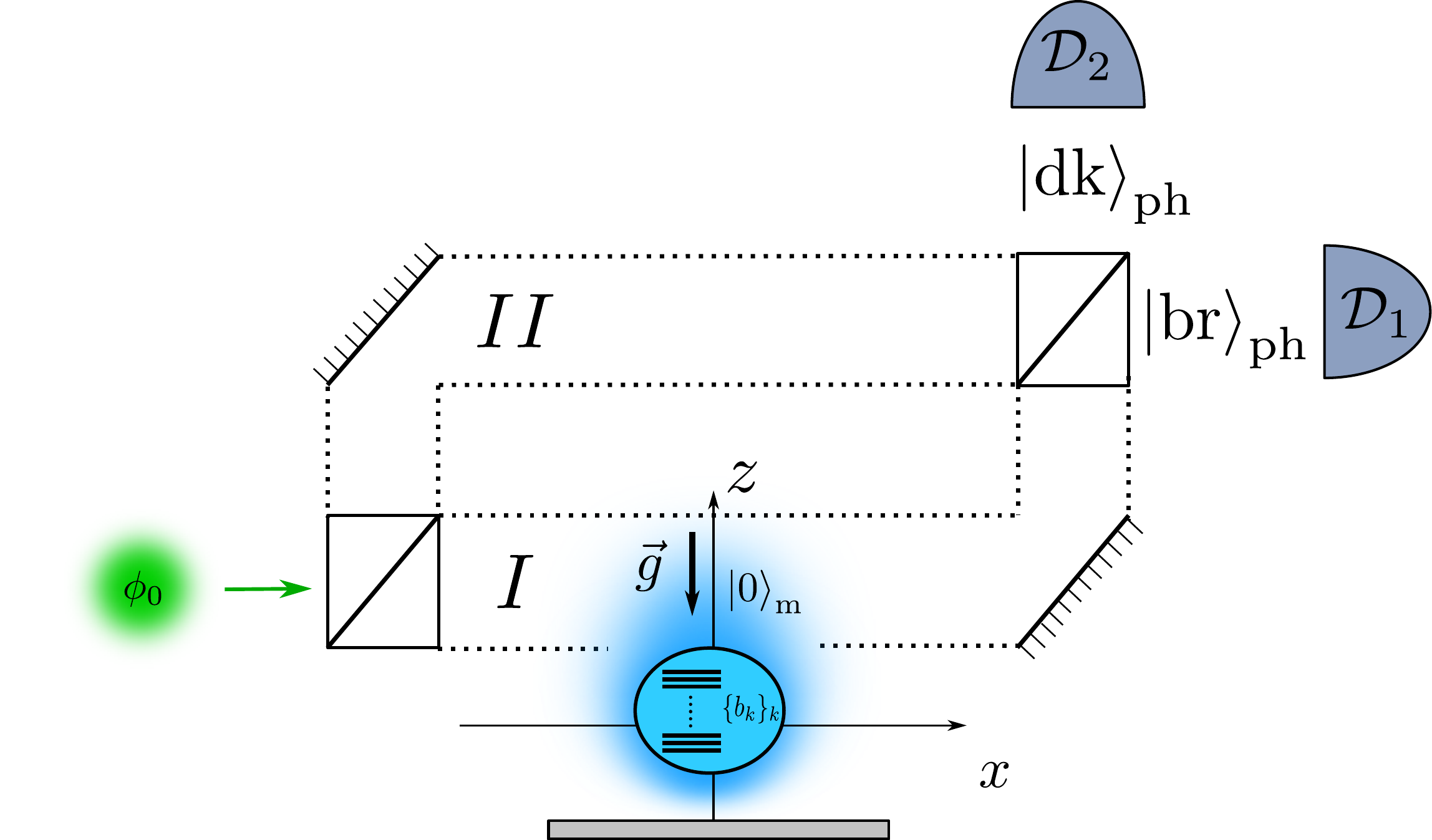}
\end{center}
\caption{Interaction-free measurement-driven engine. The bomb is prepared in its motional ground state that has some spatial overlap with the photon mode in the arm $I$ of a Mach-Zehnder interferometer. A single photon is sent through the input beamsplitter. In the absence of a bomb, the photon exits the interferometer always in the mode $\ket{\mathrm{br}}_\mathrm{ph}$ and only detector ${\cal D}_1$ clicks. If the photon goes through arm $I$ and the bomb is present, the photon is absorbed. The presence of the bomb modifies the interference pattern, and detector ${\cal D}_2$ can fire. This latter event is associated with a projection of the bomb inside the arm $I$, causing an increase of its average potential energy, compensated by a variation of the photon's frequency. One can then raise the balanced platform without doing any work. At the end of the process, work has been done against the gravity potential and is stored in the particle's gravitational potential energy. \label{f:Engine}}
\end{figure}

\section{Position detection of a bomb}\label{s:Model}

We consider a bomb described by one motional and many internal degrees of freedom. The bomb has some spatial extension along the photon's path characterized by the length $L$, in the $x$ direction (see Fig.~\ref{f:Meas}). Decomposing the bomb into elementary slices of width $dx$, we describe the internal degree of freedom in each of them by a collection of harmonic modes of Hamiltonian $H_\mathrm{b} = \sum_k \nu_k b_k^\dagger b_k$. Here $b_k$ is the bosonic annihilation operator in mode $k$.  If the bomb is present along the photon's path, we assume that the photon interacts weakly, locally with each of these elementary slices. Moreoever, we assume that the collection $H_\mathrm{b}$ has a short correlation time $\tau_c$ and therefore behaves as a zero-temperature reservoir able to absorb the photon. On the other hand, the motional degree of freedom is characterized by Hamiltonian $H_\mathrm{m} = \frac{p_\mathrm{m}^2}{2m} + V_\mathrm{m}(z_\mathrm{m})$, where $z_\mathrm{m},p_\mathrm{m}$ are the position and momentum of the bomb's center of mass along the $z$-axis and $m$ its mass. $V_\mathrm{m}(z)$ is a potential that is assumed to be infinite for $z<0$ and linearly increasing for $z>0$. This potential is realized in particular in the quantum bouncing ball problem, i.e. for a massive particle in the gravity field on top of a rigid floor. We also assume that the bomb is tightly trapped in the $x$ direction such that the corresponding dynamics can be neglected. It is useful to introduce the ground and first excited eigenstates $\{\ket{0}_\mathrm{m},\ket{1}_\mathrm{m}\}$ of $H_\mathrm{m}$ associated with energies $0$ and $\hbar\omega_\mathrm{m}$. The bomb is initially in the ground state of all degrees of freedom, denoted as $\ket{0}_\mathrm{m}\ket{0}_\mathrm{b}$.

To detect the position of the bomb, a single photon is sent with an average momentum parallel to the $x$-axis, in a mode with finite spatial overlap with the initial bomb's wavefunction. A photon-counter is positioned to detect when the photon is transmitted (see Fig.~\ref{f:Meas}). We assume that the interaction between the photon and the atom is proportional to the projector $\Pi_\mathrm{in} = \ket{\mathrm{in}}_\mathrm{m}\bra{\mathrm{in}}$, where we have introduced a state of the bomb's center of mass $\ket{\mathrm{in}}_\mathrm{m}$ located inside the photon's beam path and fulfilling together with its orthogonal state $\ket{\mathrm{out}}_\mathrm{m}$:
\bb
\ket{0}_\mathrm{m} &=& \frac{1}{\sqrt{2}}(\ket{\mathrm{in}}_\mathrm{m}+\ket{\mathrm{out}}_\mathrm{m})\label{0st}\\
\ket{1}_\mathrm{m} &=& \frac{1}{\sqrt{2}}(-\ket{\mathrm{in}}_\mathrm{m}+\ket{\mathrm{out}}_\mathrm{m})\label{1st}.
\ee
This condition is approximately verified with a good precision for the two first energy eigenstate of the quantum bouncing ball and a Gaussian photon spatial mode (see Appendix). Here, both states $\ket{\mathrm{in}}_\mathrm{m}$ and $\ket{\mathrm{out}}_\mathrm{m}$ have the same average energy $\hbar\omega_\mathrm{m}/2$. 
The initial photon state is assumed to be a wavepacket of central frequency $\omega_\mathrm{ph}$ and spectral width $\Delta \omega_\mathrm{ph}$. This state is denoted as 
\bb
\ket{I}_\mathrm{ph} \equiv \sum_k \phi_0(\omega_k)a_{I,k}^\dagger \ket{0}_\mathrm{ph},
\ee
while $\ket{0}_\mathrm{ph}$ refers to the electro-magnetic vacuum, $a_{I,k}^\dagger$ creates a photon in a mode of frequency $\omega_k$, and $\phi_0$ is the photon wavepacket in frequency representation.

The coupling Hamiltonian $V$ between the photon and each slice of the bomb induces the absorption of the photon when the atom is localized in state $\ket{\mathrm{in}}$: 
\bb
V = i \Pi_\mathrm{in}\sum_{j,k}g_{k}(a_j^\dagger b_k - b_k^\dagger a_j),\label{V}
\ee 
where $g_k$ is the coupling strength, assumed real and independent of the photon frequency for the sake of simplicity.

The interaction is switched on at time $t=0$ when the beginning of the wavepacket reaches the bomb, and switched off at time $t=\tau = L/c$ when the end of the wavepacket leaves the bomb. In the Appendix, we derive the dynamics of the photon and bomb center of the mass, given the bomb does not explode, using standard approximations for a quantum open system interacting weakly with a reservoir \cite{Breuer}. We focus on the regime: 

\bb\label{ineq}
\omega_\mathrm{ph},\Delta\omega_\mathrm{ph}\gg \omega_\mathrm{m} \gg \Gamma
.
\ee
Here $\Gamma = \sum_k g_k^2\delta(\nu_k-\omega_\mathrm{ph})$ is the decay rate of the photon. 
As it will be clearer below, the inequality $\Delta\omega_\mathrm{ph}\gg \omega_\mathrm{m}$ is required for the interaction-free measurement to work. 
We find that if the photon leaves the bomb area without exploding it, which happens with $50\%$ probability, the system is in state (see Appendix):

\bb
\ket{\tilde\Psi_1}  &=&\frac{1}{{\cal N}_1(\tau)}\sum_j \bigg[(\phi_0(\omega_j)\ket{0}_\mathrm{m}+\phi_0(\omega_j+\omega_\mathrm{m})\ket{1}_\mathrm{m})\nonumber\\
&&+  e^{-\Gamma\tau/2}(\phi_0(\omega_j)\ket{0}_\mathrm{m}-\phi_0(\omega_j+\omega_\mathrm{m})\ket{1}_\mathrm{m})\bigg]a_{I,j}^\dagger\ket{0}_\mathrm{ph}\ket{0}_\mathrm{b}.\label{psi1}
\ee
Here ${\cal N}_1(\tau) = \sqrt{2(1+e^{-\Gamma\tau})}$ is a normalization factor. The tilde indicates that the state is written in the interaction picture w.r.t. Hamiltonians $H_\mathrm{m}+H_\mathrm{ph}$. 
Now, if $\Gamma\tau\gg 1$, i.e. for a bomb that perfectly absorbs the photon when they are in contact, the second line terms vanish. In addition, condition $\omega_\mathrm{m}\ll \Delta\omega_\mathrm{ph}$ implies that the shifted wavepacket $\phi_0(\omega+\omega_\mathrm{m})$ is almost undistinguishable from the initial photon wavepacket $\phi_0(\omega)$. Consequently, state $\ket{\tilde\Psi_1}$ can be well approximated by $\ket{I}_\mathrm{ph}\ket{\mathrm{out}}_\mathrm{m}\ket{0}_\mathrm{b}$, i.e. an absence of absorption of the photon is associated with the bomb collapsed to the outside state. Note that the motional state $\ket{\mathrm{out}}_\mathrm{m}$ is not an eigenstates of $H_\mathrm{m}$ and therefore evolves coherently when written back in Schr\"odinger picture. This evolution is a coherent rotation exchanging the roles of states $\ket{\mathrm{in}}_\mathrm{m}$ and $\ket{\mathrm{out}}_\mathrm{m}$ at frequency $\omega_\mathrm{m}$, which conserves the energies of the bomb and the photon. Such unitary evolution can be corrected by letting the bomb evolve freely after the end of the interaction with the photon, during a time chosen to compensate the accumulated phase.\\

\section{Energy exchanges}\label{s:EnEch}
The coupled evolution of the bomb and the photon leads to energy exchanges between the two systems. Interestingly, even when the photon is not absorbed (the bomb is found outside the beam path), the bomb's energy still increases from its ground state due to the localization of its motional wavefunction. This measurement-induced energy increase can then be extracted to build a measurement-driven engine \cite{Elouard17Maxwell,Yi17,Elouard18Engine,Bresque20}, for instance by taking advantage of the fact that state $\ket{\mathrm{in}}_\mathrm{m}$ has no support on the region immediately near the floor to raise it without paying work. Precisely, the energy exchange is reflected in Eq.~\eqref{psi1} by the fact that motional state $\ket{0}_\mathrm{m}$ (resp. $\ket{1}_\mathrm{m}$) of the bomb is always associated with a photon wavepacket centered around frequency $\omega_\mathrm{ph}$ (resp. $\omega_\mathrm{ph}-\omega_\mathrm{m}$), ensuring energy conservation. During the process, the bomb's expected internal energy goes from $\moy{H_\mathrm{m}(0)}+\moy{H_\mathrm{b}(0)} = 0$ to 
$\bra{\mathrm{out}}_\mathrm{m}\bra{0}_\mathrm{b}(H_\mathrm{m}+H_\mathrm{b})\ket{\mathrm{out}}_\mathrm{m}\ket{0}_\mathrm{b} = \hbar\omega_\mathrm{m}/2$. This energy is provided by the photon, whose expected energy goes from ${}_\mathrm{ph}\bra{I}H_\mathrm{ph}\ket{I}_\mathrm{ph} = \hbar\omega_\mathrm{ph}$ to  $\bra{\psi_1}H_\mathrm{ph}\ket{\psi_1} = \hbar\omega_\mathrm{ph}-\hbar\omega_\mathrm{m}/2$.
 
Although the transmission of the photon is associated with the projection of the bomb outside of the photon's path, it seems far-fetched at this point (i.e without interferometer yet),  to claim that the bomb and the photon did not interact. In the Appendix, where the bomb-photon evolution is derived, we show that the localization of the bomb wavefunction and the shift in the photon frequency come from terms in second order in $V$, corresponding to virtual absorptions and re-emissions of the photon at a different frequencies. The same result was obtained in a closely analogous situation by Dicke \cite{Dicke81}. In both cases, one could argue that the two systems actually interacted, and this interaction resulted in pushing up the bomb out of the photon's path without blowing it up.

In the following, we include this atom position measurement inside of a Mach-Zehnder interferometer enabling the implementation of a quantum bomb tester as proposed by Elitzur and Vaidman \cite{Elitzur93}. 

\section{Interaction-free measurement of the bomb's position}\label{s:IntFreeEngine}

We now consider that the photon path described above is in one of the two arms of a Mach-Zehnder interferometer, labeled $I$ and $II$ (see Fig.~\ref{f:Engine}).  We assume that at $t=0$ a single photon is sent through the first beam-splitter preparing the state $(\ket{I}_\mathrm{ph}+\ket{II}_\mathrm{ph})/\sqrt{2}$. We have introduced 
\bb
\ket{II}_\mathrm{ph} = \sum_k \phi_0(\omega_k)a_{II,k}^\dagger\ket{0}_\mathrm{ph}
\ee
involving the same initial photon wavepacket as before but in arm $II$ instead of arm $I$, and $\ket{0}_\mathrm{ph}$ is now the vacuum in both arms. The bomb is prepared in its ground state as before, such that the total system initial state is 
\bb
\ket{\Psi_0'} = \frac{1}{\sqrt{2}}(\ket{I}_\mathrm{ph}+\ket{II}_\mathrm{ph})\ket{0}_\mathrm{m}\ket{0}_\mathrm{b}.
\ee
 As before, the photon and the atom interact via Hamiltonian $V$ between times $0$ and $\tau$ fixed by the bomb size, solely if the photon and the atom are both present in arm $I$. As a consequence, the evolved state at time $\tau$ can be deduced from Eq.~\eqref{psi1} by noting that the branch of $\ket{\Psi_0'}$ involving state $\ket{II}_\mathrm{ph}$ does not evolve. Assuming the same conditions as in previous section, in particular given by Eq.~\eqref{ineq}, we obtain (in the interaction picture):

\bb
\ket{\tilde\Psi_1'} &=&  \frac{1}{{\cal N}_1'(\tau)}\bigg[\ket{II}_\mathrm{ph}\ket{0}_\mathrm{m}\ket{0}_\mathrm{b}+\sum_j \bigg\{\frac{e^{-\Gamma\tau/2}}{2}\Big(\phi_0(\omega_j)\ket{0}_\mathrm{m}-\phi_0(\omega_j+\omega_\mathrm{m})\ket{1}_\mathrm{m}\Big)\nonumber\\
&&+ \frac{1}{2}\Big(\phi_0(\omega_j)\ket{0}_\mathrm{m}+\phi_0(\omega_j+\omega_\mathrm{m})\ket{1}_\mathrm{m}\Big) \bigg\}a_{I,j}^\dagger\ket{0}_\mathrm{ph}\ket{0}_\mathrm{b}\bigg]\quad\,\,\label{psi1pex}\\
&&\simeq  \frac{1}{{\cal N}_1'(\tau)}\bigg[\ket{II}_\mathrm{ph}\ket{0}_\mathrm{m}\ket{0}_\mathrm{b}+\frac{1}{\sqrt 2}\ket{I}_\mathrm{ph}\ket{\mathrm{out}}_\mathrm{m}\ket{0}_\mathrm{b}+\frac{e^{-\Gamma\tau/2}}{\sqrt 2}\ket{I}_\mathrm{ph}\ket{\mathrm{in}}_\mathrm{m}\ket{0}_\mathrm{b}\bigg],\nonumber\\\label{psi1p}
\ee
where ${\cal N}_1'(\tau) = \sqrt{3+e^{-\Gamma\tau}}/\sqrt{2}$. As in the previous section, we have used the fact that the photon state is almost unaffected as the frequency shift $\omega_\mathrm{m}$ is much smaller than its initial variance $\Delta\omega_\mathrm{ph}$ in order to give an approximate factorized form. This operating condition is crucial to preserve interference between the transmitted photon state in arm $I$ and the photon state in arm $II$ at the output of the final beam-splitter.

The two exit ports of the interferometer are monitored by photo-counters. In the absence of the bomb, the photon always goes out of the interferometer in the same mode, called bright (label br), and the other port's detector never fires: this port is called the dark port (label dk). In the presence of a bomb able to absorb the photon, the interference pattern is modified, leading to a non-zero probability for the dark port to fire. Such an event thus allows us to tell for sure that the bomb was present in arm $I$  \cite{Elitzur93}. The exit beam-splitter links the inside and outside modes according to:
\bb
\ket{I}_\mathrm{ph} &\rightarrow & \frac{1}{\sqrt 2}(\ket{\mathrm{br}}_\mathrm{ph}+\ket{\mathrm{dk}}_\mathrm{ph})\label{BSI}\\
\ket{II}_\mathrm{ph} &\rightarrow & \frac{1}{\sqrt 2}(\ket{\mathrm{br}}_\mathrm{ph}-\ket{\mathrm{dk}}_\mathrm{ph}).\label{BSII}
\ee
 After the photon exits the interferometer, the state of the system is then (in the interaction picture) $\ket{\tilde\Psi_2}\propto$
\bb
\ket{\mathrm{br}}_\mathrm{ph}\big\{(1+e^{-\Gamma\tau/2})\ket{\mathrm{in}}_\mathrm{m}+2\ket{\mathrm{out}}_\mathrm{m}\big\}\ket{0}_\mathrm{b} -(1-e^{-\Gamma\tau/2})\ket{\mathrm{dk}}_\mathrm{ph}\ket{\mathrm{in}}_\mathrm{m}\ket{0}_\mathrm{b}.\quad\label{psi2p}
\ee
\\
This allows us to conclude that the dark port fires with probability $p_\mathrm{dk} = (1-e^{-\Gamma\tau/2})^2/8$ which goes to $1/8$ in the limit $\Gamma\tau\gg 1$. Moreover, when this happens the bomb is projected inside the interferometer. Specifically, the conditioned state is $\ket{\phi_\mathrm{dk}}= \ket{\mathrm{in}}_\mathrm{m}\ket{0}_\mathrm{b}$: the bomb did not explode, which corresponds to the usual interpretation of an \emph{interaction-free measurement}, but the motional degree of freedom got collapsed inside the arm. The absence of the bomb is recovered taking $\Gamma = 0$, such that this event never occurs. Note that when the bright port fires, which occurs with probability $p_\mathrm{br} = (5+2e^{-\Gamma\tau/2}+e^{-\Gamma\tau})/8$, one cannot conclude with certainty that the bomb is located outside. In this case the bomb state is updated to $\ket{\phi_\mathrm{br}}\propto \big\{(1+e^{-\Gamma\tau/2})\ket{\mathrm{in}}_\mathrm{m}+2\ket{\mathrm{out}}_\mathrm{m}\big\}\ket{0}_\mathrm{b}$.   
Finally, the absence of detection is associated with the explosion of the bomb and the failure of the interaction-free measurement, which occurs with probability $p_\mathrm{expl} = (1-e^{-\Gamma\tau})/4$.


In the case where the detector at the dark port clicked, one can infer that (i) the bomb was present in the arm $I$ as the interference pattern was modified, and (ii) the photon was not absorbed by the bomb, otherwise it could not have been detected. If one in addition assumes a perfect absorption, i.e. $\Gamma\tau\gg 1$, one finds that the only term in $\ket{\tilde\Psi_1'}$ compatible with these conditions is the one proportional to $\ket{II}_\mathrm{ph}$: it seems natural to conclude that the photon must have taken path $II$ \cite{Elitzur93}. 

On the other hand, as explained above, the bomb is projected onto the state $\ket{\mathrm{in}}_\mathrm{m}\ket{0}_\mathrm{b}$ when this outcome is obtained, and this corresponds to an increase of its internal energy by $\hbar\omega_\mathrm{m}/2$, provided by the photon. Just as before, this energy could be extracted in a setup similar to the one described in Ref.~\cite{Elouard18Engine}. One therefore faces the following dilemma to interpret the phenomenon: Either (a) the photon was allowed to provide energy to the atom \emph{at a distance}, without passing through arm $I$, despite a strictly local interaction Hamiltonian $V$, or (b) it is impossible to interpret the photon as taking only path $II$, even though the bomb was present to absorb it in arm $I$. 

Note that when the bright port fires, the updated state of the bomb $\ket{\phi_\mathrm{br}}$ is still an excited state of energy $\hbar\omega_\mathrm{m}/10$ (for $\Gamma\tau\gg 1$), and this energy is also in that case provided by the photon (see Appendix).

\section{Insights from weak value analysis}\label{s:WeakVal}

We now analyze the situation using the weak value of observables involved of the problem. This quantity allows to focus on postselected ensembles -- here on the case where the dark port clicks and the interaction-free energy transfer indeed occurs. The weak value of a given operator $A$ on the photon-bomb Hilbert space at time $t$, is defined as:
\bb
 \moy{A}_\mathrm{w}(t) = \frac{\bra{\Psi_f}U_\mathrm{BS}{\cal U}(\tau-t)A {\cal U}(t)\ket{\Psi_i}}{\bra{\Psi_f}U_\mathrm{BS}{\cal U}(\tau)\vert\Psi_i\rangle},
 \ee
where $\ket{\Psi_i}$ (resp. $\ket{\Psi_f}$) corresponds to a state in which the system is prepared (resp. postselected). Here ${\cal U}(t)$ is the (non-unitary) propagator encoding the evolution of the photon-bomb state up to time $t$, postselecting on the absence of bomb explosion. $U_\mathrm{BS}$ is the unitary transformation describing the action of the beam-splitter, acting on $\ket{I}_\mathrm{ph}$ and $\ket{II}_\mathrm{ph}$ as described by Eq.~\eqref{BSI} -Eq.~\eqref{BSII}, and leaving $\ket{0}_\mathrm{ph}$ unchanged.

The weak value can be interpreted as the average outcome of a weak measurement of observable $A$ performed at time $t$ \cite{Aharonov1988Apr}. Such weak measurements have the advantage of having negligible back-action on the system state, at the cost of yielding very noisy outputs and therefore requiring many repetitions for the average to be resolved. Here we choose $0\leq t\leq \tau$, $\ket{\Psi_i}=\ket{\Psi_0'}$ and post-select on the dark port firing (this implies that the bomb did not explode and is projected into state $\ket{\mathrm{in}}_\mathrm{m}$), i.e. in the interaction picture $\ket{\tilde\Psi_f} = \ket{\mathrm{dk}}_\mathrm{ph}\ket{\mathrm{in}}_\mathrm{ph}\ket{0}_\mathrm{b}$. The weak value is plotted in Fig.~\ref{f:WeakValues}a for $A$ being different projectors. The weak values evoked below are computed in the Appendix for the current model. 

\begin{figure}[h!]
\begin{center}
\includegraphics[width=0.7\textwidth]{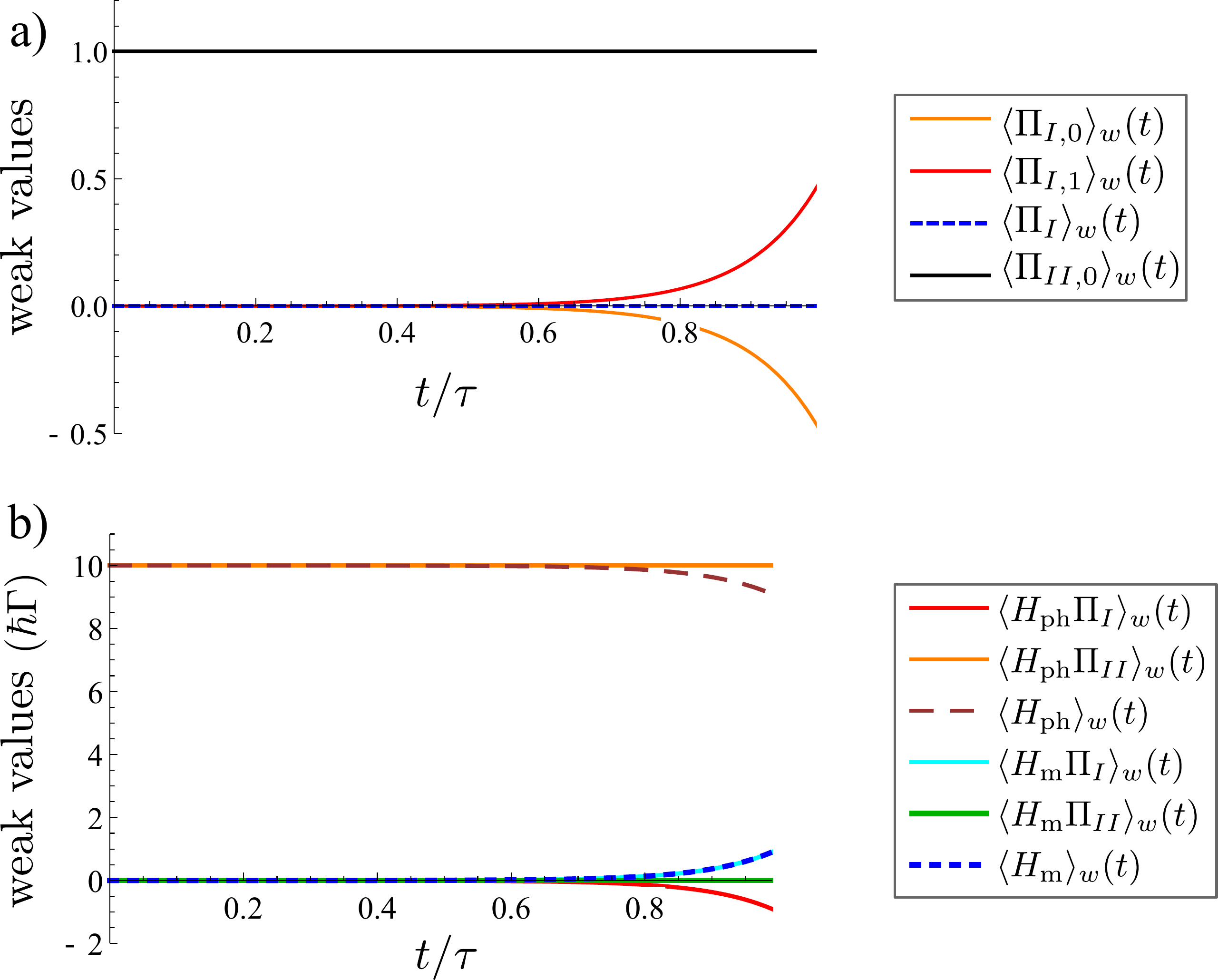}
\end{center}
\caption{Weak values $\moy{A}_\mathrm{w}(t)$ as a function of time $0\leq t\leq\tau$ of different observables $A$ involved in the problem. a) Projectors on the photon in arm $I$  and the bomb in the ground motional state (orange), photon in arm $I$  and bomb in the excited motional state (red), photon in arm $I$  (dashed blue), photon in arm $II$ and bomb in the ground motional state (black). b) Energy of the photon in arm $I$  (red), in arm $II$ (orange), total photon energy (dashed brown), energy of the motional degree of freedom of the bomb when the photon is in arm $I$  (cyan), in arm $II$ (green) and total motional energy (dotted blue). Hamiltonian weak values are plotted in units of $\hbar\Gamma$. Parameters: $\Gamma\tau =20$, $\omega_\mathrm{m}/\Gamma = 2$, $\omega_\mathrm{ph}/\Gamma = 10$. \label{f:WeakValues}}
\end{figure}

We first note that the weak value of the projector onto arm $II$, $\Pi_{II} = \ket{II}\bra{II}$ is one at any time (black solid line in Fig.~\ref{f:WeakValues}a) while the weak value of the projector onto arm $I$, $\Pi_{I} = \ket{I}\bra{I}$, is always zero (blue dashed line in Fig.~\ref{f:WeakValues}a), supporting the interpretation that the photon solely took path $II$. We moreover compute the weak values of joint photon-bomb observables involving a projector of the photon onto one of the arms and a projector of the bomb onto state $\ket{\mathrm{in}}_\mathrm{m}$ or $\ket{\mathrm{out}}_\mathrm{m}$. One finds that for $\Gamma\tau\gg 1$, solely the weak value of the projector $\Pi_{II,\mathrm{in}}$, i.e. of the photon being in arm $II$ while the bomb is projected inside, in non zero and remains $1$. This result is therefore compatible with an energy transfer occurring at a distance, despite a local interaction Hamiltonian. 

Some insights about how this exchange is mediated can be obtained if one considers weak values involving the projectors onto another basis for the bomb, namely the motional energy eigenbasis $\{\ket{0}_\mathrm{m},\ket{1}_\mathrm{m}\}$ instead of the $\{\ket{\mathrm{in}}_\mathrm{m},\ket{\mathrm{out}}_\mathrm{m}\}$. We find that the weak values of projectors onto states $\ket{I}_\mathrm{ph}\ket{0}_\mathrm{m}$ and $\ket{I}_\mathrm{ph}\ket{1}_\mathrm{m}$ actually acquire non-zero values during the interaction (see Fig.~\ref{f:WeakValues}a): both start at zero and become finite during the photon-bomb joint evolution. Strikingly, the weak value $\moy{\Pi_{I,0}}_\mathrm{w}(t)$ of the photon being in arm $I$ and the bomb in its motional ground state is negative, while the weak value for the photon being in arm $I$ and the bomb being in the excited ground state is always positive and fulfills $\moy{\Pi_{I,1}}(t) = -\moy{\Pi_{I,0}}(t)$. We therefore recover by tracing over the bomb subspace (wich simply amounts to summing $\moy{\Pi_{I,0}}(t)$ and $\moy{\Pi_{I,1}}(t)$) that the weak value of the photon projector on arm $I$ is zero. The existence of these two non-zero values implies that even if the average excitation number in the photonic modes in arm $I$ is zero, the quantum correlations building up between the photon and bomb motional state behave as if the bomb was emitting a pair of eletromagnetic excitations, one of them effectively of negative population. We investigate further this phenomenon by studying the Hamiltonian weak values, plotted in Fig.~\ref{f:WeakValues}b. We find that as expected, the energy of the photon decreases by an amount $\hbar\omega_\mathrm{m}/2$, which is in turn provided to the bomb. But, strikingly, the weak value $\moy{H_\mathrm{ph}\Pi_{II}}_\mathrm{w}(t)$ of the energy of the photonic modes in arm $II$ stays constant and equal to the energy of the incoming photon $\hbar\omega_\mathrm{ph}$. It is actually the energy of photonic modes in arm $I$ which decrease from its initial value $0$ to the negative value $-\hbar\omega_\mathrm{m}/2$ (see red solid curve in Fig.~\ref{f:WeakValues}b). These values suggest that the energy exchange between the bomb and the photon can be interpreted as taking place in two steps: First, it is mediated by an effective ``energy hole'' or a ``negative-energy photon'', which is created in arm $I$ during the interaction (remember that all the weak values associated with the projector on arm $I$ are zero at $t=0$ and build up during the interaction). Such an effective particle then propagates until the second beam-splitter where it coalesces with the photon amplitude coming from arm $II$. 

This last interpretation must evidently be taken with caution, as it involves \emph{anomalous} weak values, i.e. a values out of the observable's eigenvalues range: this is the case of $\moy{\Pi_{I,0}}(\tau)$ and $\moy{H_\mathrm{ph}\Pi_{I}}_\mathrm{w}(\tau)$ which are both negative. Such anomaly has been proved to be a hallmark of the presence of contextuality \cite{Pusey2014Nov}, and is equivalent to the violation of a generalized Leggett-Garg inequality \cite{Leggett1985Mar,Williams2008Jan,Palacios-Laloy10,Goggin11,Suzuki12,Groen13}. This violation precisely demonstrates the incompatibility of quantum mechanics with the hypotheses of \emph{macroscopic realism} and \emph{nonivasive measurements} -- namely that observables take definite values at any time and that the latter can be determined with arbitrarily small disturbance of the system. The apparent contradiction between the absence of the photon in arm $I$ and the energy exchange with the bomb is another example of quantum effect that defy classical logic, such as the violation of the pigeon hole principle \cite{Aharonov16,Chen19}, or the Quantum Cheshire Cat paradox \cite{Aharonov13}. The phenomenon reported in this article shares some similarities with the latter paradox, in which a single photon seemingly takes one arm of an interferometer, while still having some of its properties, like its polarization, be present in the other arm. Just as in our case, the phenomenon can be diagnosed through weak values summing up to zero when tracing over the subspace corresponding to this delocalized property -- the ``grin'' of the Cheshire cat. In our case, this is an energy exchange that is happening where the photon is effectively not present.

\section{Discussion and conclusion}
\label{s:Discussion}

An important requirement for the interaction-free measurement presented here to occur is that the atom remains in a coherent superposition of motional energy eigenstates after interaction with the photon and until the dark port detector fires. To emphasize this point, let us assume that a projective measurement in the $\{\ket{0}_\mathrm{m},\ket{1}_\mathrm{m}\}$ basis is performed after the interaction, i.e. on state $\ket{\Psi_1'}$. If state $\ket{1}_\mathrm{m}$ is found, the system collapses onto state $\ket{I}_\mathrm{ph}\ket{1}_\mathrm{m}\ket{0}_\mathrm{b}$.
As this state has no amplitude on arm $II$, it leads to no interference at the second beam-splitter: in other words the energy measurement on the atom brought which-path information. Incidentally, the energy exchange between the atom and the photon is in this case localized inside arm $I$ where both systems are present. Thus, one can conclude that the seeming non-locality of the energy exchange is closely related to the uncertainty about the atom's precise energy in the final state $\ket{\mathrm{in}}_\mathrm{m}$.  On the other hand, if one lets the photon exit the second beam-splitter and be detected at the dark port, and then performs a projective measurement in the $\{\ket{\mathrm{in}}_\mathrm{m},\ket{\mathrm{out}}_\mathrm{m}\}$ basis (up to the free rotation induced by $H_\mathrm{m}$), one obtains the result $\ket{\mathrm{in}}_\mathrm{m}$ with certainty, and one can in principle extract deterministically all the energy stored in this state by doing a unitary rotation back to the ground state \cite{Elouard18Engine}. In this case, the interference between the two paths does take place: the which-path information is not directly available and can only be inferred indirectly from the fact that the photon was not absorbed. At the same time, the measurement outcome brings no information about the atom's motional energy (state $\ket{\mathrm{in}}_\mathrm{m}$ has maximum uncertainty about the energy of this degree of freedom). 

As mentioned earlier, the interaction-free measurement also requires a large energy uncertainty of the photon state $\hbar\Delta\omega_\mathrm{ph} \gg \hbar\omega_\mathrm{m}$. As discussed in \cite{Vaidman2003Mar}, one can assess the interaction-free nature of the measurement by looking at how the post-selected wavefunction, i.e. $\ket{\tilde\Psi_\mathrm{f}}=\ket{\mathrm{dk}}_\mathrm{ph}\ket{\mathrm{in}}_\mathrm{m}\ket{0}_\mathrm{b}$ evolves backwards in time. The evidence that the photon could not interact with the bomb is then deduced from the fact that this backward wavefunction is zero in the arm containing the bomb. Here, we find that the wavefunction of the backward propagated wavefunction in arm $I$ but scales at most like $\omega_\mathrm{m}/\Delta\omega$, such that an effective absence of interaction is realized for large enough initial photon frequency variance. This constraint can also be understood as a fundamental upper bound on the amount of energy the photon can exchange with the bomb without blowing it, which has to be much smaller than $\hbar\Delta\omega$.

Implementing this setup in order to test the results of this article seems challenging. Indeed, the typical position uncertainty of single atoms in the gravity field is very small and an ideally absorbing bomb requires either a strong coupling or a large spatial extend of the bomb. However, state-of-the-art setups on photonic crystal waveguides \cite{Douglas2015Apr} enable the engineering of strong coupling between single-photon and atoms. In such a setup, the role of the ``inside'' and ``outside'' states could correspond to two different modes of the atom coupled differently to the photonic modes. In addition, it is not actually essential to have an absorption process to model the explosion of the bomb: a scattering process which would divert the photon out of interferometer is also perfectly suitable and was used in experimental implementations of Elitzur-Vaidman bomb tester \cite{Kwiat95}. An ideal bomb then only needs to have the sum of the scattering and absorption probabilities to be close to one when it is inside the arm. This simplification allows another implementation exploiting the progresses of opto-mechanics that allowed to cool macroscopic oscillators down to the ground state \cite{Teufel11,Chan11} and prepare them into non-classical state \cite{O'Connell2010Mar}. In this case, the role of the bomb could be played by a movable mirror mounted on a torsion pendulum that would replace the mirror of arm $I$ on Fig.~\ref{f:Engine}. The ground state of the pendulum has an intrinsic uncertainty about the torsion angle, such that the reflected photon could be efficiently directed to the output beamsplitter, or conversely diverted out of the interferometer arm. The latter event would play the role of the bomb explosion. Interestingly, for long enough interferometer arms, a small angle uncertainty could result in a large position uncertainty of the outgoing photon, amplifying the phenomenon.

Finally, we emphasize that the present effect relies on the interaction between two degrees of freedom prepared in coherent superpositions, a situation which is also at the core of Hardy's paradox~\cite{Hardy92} involving two intertwined interferometers. The present setup differs from this earlier work in two ways: (i) the coherent superposition of one of the two degrees of freedom involve states of different energies, (ii) one of the two degrees of freedom (the bomb position) is continuous. Contrary to the first one, the second difference is not actually crucial to observe the energy exchange. The possibility to involve a continuous degree of freedom in an interaction-free measurement was demonstrated in Ref.~\cite{Rogers20}. 

We have analyzed an interaction-free measurement setup in a situation where the measured bomb can gain energy, even though it seemingly did not interact with the meter, a single photon. This energy gain can be understood as a non-local energy transfer triggered by a strictly local interaction Hamiltonian. 
Weak value analysis postselecting on the cases where the interaction-free measurement successfully happens shows that the probability to find the photon in the arm of the bomb indeed vanishes. However, it also reveals the contextual nature of the quantum correlations between the photon and the bomb which still allow for an energy transfer to occur. This transfer can indeed be tracked in the energy-valued weak-values, some of them taking anomalous (negative) values. Regardless of interpretation, this interaction free quantum measurement engine is able to lift the most sensitive bomb without setting it off. These results bring new insights about quantum energy transfers, opening new path to design genuinely quantum engines.\\

\begin{acknowledgements}
Work by CE and ANJ was supported by the US Department of Energy (DOE), Office of Science, Basic Energy Sciences (BES), under Grant No. DE-SC0017890. M.W. acknowledges the Fetzer Franklin Fund of the John E. Fetzer Memorial Trust. This research was partly conducted at the KITP, a facility supported by the National Science Foundation under Grant No. NSF PHY-1748958.
\end{acknowledgements}

\section*{Appendix A: Inside and outside states}

The quantum bouncing ball Hamiltonian $H_\mathrm{m}$ corresponds to a potential $V_\mathrm{m}(z) = mgz$ for $z>0$, and infinite for $z\leq 0$. The wavefunction of the energy eigenstates  of $H_\mathrm{m}$, i.e. $\ket{j}_\mathrm{m}$ with $j$ an integer, can be expressed given in term of the Airy function ${\cal A}(u)$ which is the solution of the ODE $\frac{d^2}{du^2}y(u)-u y(u)=0$ which does not diverge for $u\to\infty$:
\bb
\psi_\mathrm{m}^{(j)}(z) = \langle z \ket{j}_\mathrm{m} = \frac{{\cal A}(z/z_0+\zeta_{j+1})}{{\cal A}'(\zeta_{j+1})},
\ee
where $z_0 = (\hbar^2/2m^2g)^{1/3}$ is a characteristic length of the problem and $\zeta_j$ is the $j$th zero of the Airy function (which is negative). 
We consider that the atom is in the ground state $\ket{0}_\mathrm{m}$ and one sends a photon towards it. The photon spatial wavefunction has a Gaussian shape in the $z$ direction $\phi_0(z)$. The photon and the atom interaction strength is proportional to the spatial overlap of their wavefunctions, such that $\phi_0(z)$ can be seen as the atom wavefunction that optimally interacts with the photon. It turns out that a Gaussian wavefunction $\chi(z)$ of average $\bar z = 2.80 z_0$ and standard deviation $\bar\sigma =  1.27x_0$ coincides almost exactly with that of the state $\ket{-}_\mathrm{m} =(\ket{0}_\mathrm{m}-\ket{1}_\mathrm{m})/\sqrt{2}$, as quantified by an overlap probability $(\int dz\chi(z)\langle z \ket{-}_\mathrm{m})^2 \simeq 0.99$. This justifies Eqs.(1)-(2) of the main text defining states $\ket{\mathrm{in}}$ and $\ket{\mathrm{out}}$. While a similar analysis can be done for an arbitrary superposition of the states $\ket{0}_\mathrm{m}$ and $\ket{1}_\mathrm{m}$ being coupled to the photon, this particular choice simplifies the calculation and clarifies the analysis.\\

As an interesting additional feature, the wavefunction of state $\ket{\mathrm{in}}_\mathrm{m}$ turns out to be negligible in the vicinity of the platform up to $z\sim z_0/2$. As a consequence, once the bomb has been projected onto this state, it is possible to raise the floor level of an amount $\Delta z \leq z_0/2$ without paying work to raise the bomb, therefore storing usefull gravitational potential energy \cite{Elouard18Engine}.

\section*{Appendix B: Photon-bomb interaction}
The bomb is a zero temperature reservoir, modeled by a collection of harmonic modes $b_k$ at frequency $\nu_k$. The photon modes are denoted $a_j$ with frequencies $\omega_j$. The coupling Hamiltonian reads:
\bb
V = i \Pi_\mathrm{in}\sum_{j,k}g_{k}(a_j^\dagger b_k - b_k^\dagger a_j).
\ee
We consider the weak coupling limit $\vert g_{k}\vert\tau_c\ll 1$ where $\tau_c$ it the correlation time of the bomb and assume a coupling independent of the photon's frequency for simplicity. We model the evolution the following way: the photon and the bomb interact during $\Delta t$ fulfilling $\tau_c\ll\Delta t\ll \tau$, then the number of excitations in the bomb is checked. In addition, we work in the limit:
\bb
\omega_\mathrm{ph}^{-1},\Delta\omega_\mathrm{ph}^{-1},\omega_\mathrm{b}^{-1},\omega_\mathrm{m}^{-1} \ll \Delta t \ll \Gamma^{-1}
\ee
where 
\bb
\Gamma = \sum_{k} g_{k}^2 \delta(\omega_\mathrm{ph}-\omega_k),\label{Gamma}
\ee
is the effective rate of the evolution induced by the interaction, as shown below. 

We work in the interaction picture (denoted with tilde) where the coupling Hamiltonian fulfils:
\bb
\tilde V(t) = i \tilde \Pi_\mathrm{in}(t)\sum_{j,k}g_{k}(a_j^\dagger b_k e^{i (\omega_j-\nu_k) t} - b_k^\dagger a_je^{i(\nu_k-\omega_j)t}).\nonumber\\
\ee
Starting from the state $\ket{\tilde\Psi(t)} = \ket{\tilde\phi(t)}_\mathrm{ph}\ket{\tilde\chi(t)}_\mathrm{m}\ket{0}_\mathrm{b}$ and the bomb in the vacuum $\ket{0}_\mathrm{b}$, the evolved state, after the bomb is found containing $n$ excitations, reads at second order in $\Delta t$:
\bb
&&\ket{\tilde\Psi(t+\Delta t)} = \delta_{n,0}\ket{\tilde\Psi(t)}-i\int_{t}^{t+\Delta t}dt' {}_\mathrm{b}\bra{n}\tilde V(t')\ket{0}_\mathrm{b}\ket{\tilde\Psi(t)}\nonumber\\
&&\quad\quad- \frac{1}{2}\int_{t}^{t+\Delta t}dt'\int_{t}^{t'}dt''{}_\mathrm{b}\bra{n}\tilde V(t')\tilde V(t'')\ket{0}_\mathrm{b}\ket{\tilde\Psi(t)}.
\ee
We focus to the case $n=0$ (bomb not exploded), such that the term in first order in $\tilde V$ vanishes. Denoting $\tilde M_0$ the Kraus operator updating the wavefunction in this case, we get:
\bb
\tilde M_0 &=&  1- \frac{1}{2}\int_{t}^{t+\Delta t}dt'\int_{t}^{t'}dt''{}_\mathrm{b}\bra{n}\tilde V(t')\tilde V(t'')\ket{0}_\mathrm{b}\nonumber\\
&=&  1+ \frac{1}{2}\int_{t}^{t+\Delta t}\!\!\!\!\!\!\!\!dt'\int_{0}^{t'-t}\!\!\!\!\!\!d\tau e^{i(\omega_i-\omega_k)t'}e^{i(\omega_k-\omega_j)(t'-\tau)}\nonumber\\
&&\quad\quad\quad\quad\times\sum_{i,j,k}g_{k}^2 a_i^\dagger a_j \tilde \Pi_\mathrm{in}(t')\tilde \Pi_\mathrm{in}(t'-\tau).
\ee
We now use that $\tilde\Pi_\mathrm{in}(t) = (1/2)(\idop - \sigma_\mathrm{m}e^{-i\omega_\mathrm{m}t}- \sigma_\mathrm{m}^\dagger e^{i\omega_\mathrm{m}t})$ and use that the bomb correlation time is assumed to be much smaller than $\Delta t$. This allows us to replace the upper bound of the integral over $\tau$ by $+\infty$. This integral can then be computed and yields Dirac distributions, forming the bomb spectral density taken at various frequencies, i.e. $S(\omega) = \sum_k g_k^2\delta(\omega-\omega_k)$, where $\omega$ takes typical values in the range $[\omega_\mathrm{ph}-\Delta\omega_\mathrm{ph},\omega_\mathrm{ph}+\Delta\omega_\mathrm{ph}]$, i.e. the frequencies typically contained in the initial photon state. We assume that the bomb spectral density is flat on this frequency range, such that we can replace $S(\omega)$ with $\Gamma$ defined in Eq.~\eqref{Gamma}. Finally, apply the Secular approximation \cite{CCT}, i.e. we neglect in the integral over $t'$ all the term rotating at non-zero frequency. We finally get (back in Scr\"odinger picture):
\bb
&&M_0 -\idop = -i(H_\mathrm{m}+H_\mathrm{ph})dt- \frac{\Gamma}{4}\sum_{ij}a_i^\dagger a_j\bigg[\delta(\omega_i-\omega_j)\nonumber\\
 &&\quad\quad-\sigma_\mathrm{m}\delta(\omega_i-\omega_j-\omega_\mathrm{m})-\sigma_\mathrm{m}^\dagger\delta(\omega_i-\omega_j+\omega_\mathrm{m}))\bigg]
\ee

This Kraus operator solely couples states $\ket{\omega_j}_\mathrm{ph}\ket{0}_\mathrm{m}\ket{0}_\mathrm{b}$ to $\ket{\omega_j-\omega_\mathrm{m}}_\mathrm{ph}\ket{1}_\mathrm{m}\ket{0}_\mathrm{b}$. As a consequence, if one assumes the ansatz,
\bb
\ket{\tilde\Psi(\tau)} &=& \sum_j\bigg[ c_0(\tau)\phi_0(\omega_j)\ket{0}_\mathrm{m}\nonumber\\
 && +  c_1(\tau)\phi_0(\omega_j+\omega_\mathrm{m})\ket{1}_\mathrm{m}\bigg]a_{I,j}^\dagger\ket{0}_\mathrm{ph}\ket{0}_\mathrm{b},
\ee
where $\phi_0(\omega)$ is the initial photon wavefunction, one finds that the amplitudes $c_0$ and $c_1$ fulfill the evolution equations:
\bb
\dot c_0 &=& -\frac{\Gamma}4 (c_0-c_1)\nonumber\\
\dot c_1 &=& \frac{\Gamma}4  (c_0-c_1).
\ee
In particular, starting from $c_0 = 1$, $c_0=0$, one gets $c_{0,1}(\tau) = (1\pm e^{-\Gamma \tau/2})/2$. Finally, the state $\ket{I}_\mathrm{ph}\ket{0}_\mathrm{m}\ket{0}_\mathrm{b}$ is mapped onto state $\sum_j(\phi_0(\omega_j)\ket{0}_\mathrm{m}+\phi_0(\omega_j+\omega_\mathrm{m})e^{-i\omega_\mathrm{m}\tau}\ket{1}_\mathrm{m})\ket{0}_\mathrm{b}/\sqrt{2}$ at long times $\tau\gg \Gamma^{-1}$. This state can be approximated (after being renormalized) by $\ket{I}_\mathrm{ph}\ket{\mathrm{out}}_\mathrm{m}\ket{0}_\mathrm{b}$ provided the two wavefunctions $\phi_0(\omega)$ and $\phi_0(\omega+\omega_\mathrm{m})$ have overlap of almost unity, i.e. provided that the initial photon width $\Delta \omega_\mathrm{ph}$ is much larger than $\omega_\mathrm{m}$. 

Note that the bomb-photon interaction time $\tau$ is set by the longest of the bomb length and photon duration. The photon duration is constrained by its frequency width $\tau_\mathrm{ph}= \Delta \omega_\mathrm{ph}^{-1} \ll \omega_\mathrm{m}^{-1}\ll \Gamma^{-1}$. Consequently, in order to have $\Gamma\tau > 1$, we need a large bomb of width $L > c/\Gamma$. 

Note also that when written in Schr\"odinger picture, the system's state for $\Gamma\tau\gg 1$ actually follows a limiting cycle $\ket{I(\tau)}\ket{\mathrm{out}(\tau)}_\mathrm{m}$. The time-dependence of the photon state wavefunction solely encode the free propagation of the wavepacket. On the other hand, the bomb motional state evolves $\ket{\mathrm{out}(\tau)}_\mathrm{m} =(\ket{0}_\mathrm{m}+e^{-i\omega_\mathrm{m}\tau}\ket{1}_\mathrm{m})/\sqrt 2$, i.e. a coherent rotation exchanging states $\ket{\mathrm{out}}_\mathrm{m}$ and $\ket{\mathrm{in}}_\mathrm{m}$ at frequency $\omega_\mathrm{m}$. In order to obtain state at the end of the protocol $\ket{\mathrm{out}}_\mathrm{m}$, one has to keep track of the phase $\omega_\mathrm{m}$ accumulated during $\tau$ to correct it by letting the bomb evolve freely during a time $\tau_2$ such that $\tau+\tau_2$ is a multiple of $2\pi/\omega_\mathrm{m}$.\\

The probability $p_\mathrm{ne}(\tau)$ of the bomb not having exploded until time $\tau$ is encoded in the norm of $\ket{\tilde\Psi(\tau)}$. We find:
\bb
p_\mathrm{ne}(\tau) = \frac{1}{2}\left(1+e^{-\Gamma\tau}\right).\label{pnetau}
\ee

\textit{Interferometer setup} -- When the previous setup is embedded as one of the two arms of an interferometer, the model is modified as follows. The coupling Hamiltonian $V$ is assumed to vanish on the photon subspace corresponding to arm $II$, i.e. the space spanned by $\{a_j^\dagger\ket{0}_\mathrm{ph}\}_j$. As the consequence, the evolution of the initial state $\ket{\Psi_0'}$ in the interaction picture can be deduced by keeping the term involving state $\ket{II}_\mathrm{ph}$ unchanged and applying to the term involving $\ket{I}_\mathrm{ph}$ the same evolution as above. The probability of non-explosion in this case can be deduced from Eq.~\eqref{pnetau} which can be understood as the conditional probability for the bomb not exploding given the photon is initially in arm $I$. Using that the probability of explosion is zero if the photon is in arm $II$, we obtain:
\bb
p_\mathrm{ne}(\tau) = \frac{1}{2} + \frac{1}{4}\left(1+e^{-\Gamma\tau}\right).
\ee
The probability of explosion is therefore $p_\mathrm{expl}(\tau) = 1-p_\mathrm{ne}(\tau) =  \frac{1}{4}\left(1-e^{-\Gamma\tau}\right)$. In order to compute the probability of the dark and bright port detections, we use the conditional probabilities 
\bb
p(\mathrm{br}\vert \mathrm{ne}) &=& \vert{}_\mathrm{m}\bra{\mathrm{br}} \tilde\Psi_2'(\tau)\rangle\vert^2\nonumber\\
&=& \frac{5+2e^{-\Gamma\tau/2}+e^{-\Gamma\tau}}{2(3+e^{-\Gamma\tau})}\\
p(\mathrm{dk}\vert \mathrm{ne}) &=& \vert{}_\mathrm{m}\bra{\mathrm{dk}} \tilde\Psi_2'(\tau)\rangle\vert^2\nonumber\\
&=&\frac{(1-e^{-\Gamma\tau/2})^2}{2(3+e^{-\Gamma\tau})},
\ee
that the photon is found in the bright and dark port respectively, given the bomb did not explode. The probabilities of these two events is then obtained by multiplying by $p_\mathrm{ne}(\tau)$, allowing to find the expressions given in the main text.

\textit{Transmitted photon energy} -- At the end of the interaction, the photon and bomb are in an entangled state as described by Eq.~(6) of the main text. One can compute the energy in the photon going into each port using the beam-splitter relation:
\bb
a^\dagger_{I,j} &=& \frac{1}{\sqrt 2}(a^\dagger_{\mathrm{br},j}+a^\dagger_{\mathrm{dk},j})\nonumber\\
a^\dagger_{II,j}&=&\frac{1}{\sqrt 2}(a^\dagger_{\mathrm{br},j}-a^\dagger_{\mathrm{dk},j}),
\ee
where $a^\dagger_{\mathrm{br},j}$ (resp. $a^\dagger_{\mathrm{dk},j}$) creates a photon of frequency $\omega_j$ at the bright (resp. dark) port. This allows us to express the exact form of the output state, introducing the notation $\phi_1(\omega_j)=\phi_0(\omega_j+\omega_\mathrm{m})$, namely $\ket{\Psi'_2} \propto$

\bb
&&\sum_j\bigg[ \Big(\phi_0(\omega_j)\frac{3+e^{-\frac{\Gamma\tau}{2}}}{2}\ket{0}_\mathrm{m}+ \phi_1(\omega_j)\frac{1-e^{-\frac{\Gamma\tau}{2}}}{2}\ket{1}_\mathrm{m}\Big) a^\dagger_{\mathrm{br},j}\nonumber\\
 &&+\frac{1-e^{-\frac{\Gamma\tau}{2}}}{2}\Big(-\phi_0(\omega_j)\ket{0}_\mathrm{m}+ \phi_1(\omega_j)\ket{1}_\mathrm{m}\Big)a^\dagger_{\mathrm{dk},j}\bigg]\ket{0}_\mathrm{ph}\ket{0}_\mathrm{b}.\nonumber\\\label{psi2pex}
\ee

The second (resp. third) line terms in Eq.~\eqref{psi2pex} can be identified (after renormalization) as the joint photon-bomb state $\ket{\Psi'_\mathrm{br}}$ (resp. $\ket{\Psi'_\mathrm{dk}}$) when the photon is found in the bright (resp. dark) port. This enables to compute the corresponding photon energy:
\bb
E_\mathrm{br}^\mathrm{(ph)} &=& \sum_j \hbar\omega_j \bra{\Psi'_\mathrm{br}} a^\dagger_{\mathrm{br},j}a_{\mathrm{br},j}\ket{\Psi'_\mathrm{br}}\nonumber\\
&=& \hbar\omega_\mathrm{ph}-\frac{\hbar\omega_\mathrm{m}}{10}\nonumber\\
E_\mathrm{dk}^\mathrm{(ph)}  &=& \sum_j \hbar\omega_j \bra{\Psi'_\mathrm{dk}} a^\dagger_{\mathrm{dk},j}a_{\mathrm{dk},j}\ket{\Psi'_\mathrm{dk}}\nonumber\\
&=& \hbar\omega_\mathrm{ph}-\frac{\hbar\omega_\mathrm{m}}{2}.
\ee
One can check that the variation of the energy of the photon from its initial energy $\hbar\omega_\mathrm{ph}$ compensates in each case the energy gained by the internal degree of freedom of the bomb:

\bb
E_\mathrm{br}^\mathrm{(m)} &=& \hbar\omega_\mathrm{m} \vert{}_\mathrm{m}\langle 1  \ket{\Psi'_\mathrm{br}}\vert^2\nonumber\\
&=& \frac{\hbar\omega_\mathrm{m}}{10}\nonumber\\
E_\mathrm{dk}^\mathrm{(m)}  &=& \hbar\omega_\mathrm{m} \vert{}_\mathrm{m}\langle 1 \ket{\Psi'_\mathrm{dk}}\vert^2\nonumber\\
&=& \frac{\hbar\omega_\mathrm{m}}{2}.
\ee

Eq.~(10) of the main text is retrieved by making the approximation $\phi_1(\omega_j)\simeq\phi_0(\omega_j)$, i.e. neglecting the photon frequency shift with respect to its much larger initial frequency uncertainty. The same approximation allows us to find that $\ket{\Psi'_\mathrm{br}} \simeq \ket{\mathrm{br}}_\mathrm{ph}\ket{\phi_\mathrm{br}}$ and $\ket{\Psi'_\mathrm{dk}} \simeq \ket{\mathrm{dk}}_\mathrm{ph}\ket{\phi_\mathrm{dk}}$.

\section*{Appendic C: Weak values} 

The evolution of the photon-atom state up to time $t$, postselecting on the absence of explosion during the interval $[0,t]$, can be encoded in the propagator:
\bb
{\cal U}(t) = \Pi_{II} + \Pi_\mathrm{i} + \Pi_\mathrm{o}e^{-\Gamma t/2},
\ee
where $\Pi_{II} = \sum_j a^\dagger_{II,j}\ket{0}_\mathrm{ph}\bra{0}a_j$ is the projector on arm $II$, and $\Pi_\mathrm{i}$ (resp. $\Pi_\mathrm{o}$) is the projector on state $\ket{\Phi_\mathrm{i}}$ (resp. $\ket{\Phi_\mathrm{o}}$) defined by:
\bb
\ket{\Phi_\mathrm{i}} &=& \sum_j\frac{\phi_0(\omega_j)\ket{0}_\mathrm{m}- \phi_0(\omega_j+\omega_\mathrm{m})\ket{1}_\mathrm{m}}{2}a_{I,j}^\dagger\ket{0}_\mathrm{ph}\ket{0}_\mathrm{b}\nonumber\\
\ket{\Phi_\mathrm{o}} &=& \sum_j\frac{\phi_0(\omega_j)\ket{0}_\mathrm{m}+ \phi_0(\omega_j+\omega_\mathrm{m})\ket{1}_\mathrm{m}}{2}a_{I,j}^\dagger\ket{0}_\mathrm{ph}\ket{0}_\mathrm{b}.\nonumber\\
\ee

We first compute the weak values associated with the rank 1 projectors $\Pi_{I,\mathrm{in}} =\ket{I}_\mathrm{ph}\bra{I}_\mathrm{ph}\ket{\mathrm{in}}_\mathrm{m}\bra{\mathrm{in}}$, $\Pi_{I,\mathrm{out}} =\ket{I}_\mathrm{ph}\bra{I}_\mathrm{ph}\ket{\mathrm{out}}_\mathrm{m}\bra{\mathrm{out}}$, $\Pi_{I,0} =\ket{I}_\mathrm{ph}\bra{I}_\mathrm{ph}\ket{0}_\mathrm{m}\bra{0}$ and  $\Pi_{I,1} =\ket{I}_\mathrm{ph}\bra{I}_\mathrm{ph}\ket{1}_\mathrm{m}\bra{1}$ (and similar for arm II). We obtain:
\bb
\moy{\Pi_{I,\mathrm{in}}}_w(t) &=& -\frac{1}{e^{\Gamma\tau}-1}\nonumber\\
\moy{\Pi_{I,\mathrm{out}}}_w(t) &=& 0\nonumber\\
\moy{\Pi_{II,\mathrm{in}}}_w(t) &=& 1+\frac{1}{e^{\Gamma\tau}-1}\nonumber\\
\moy{\Pi_{II,\mathrm{out}}}_w(t) &=& 0\nonumber\\
\moy{\Pi_{I,0}}_w(t) &=& -\frac{e^{-\Gamma (\tau-t)/2} +e^{-\Gamma \tau/2}}{2(1-e^{-\Gamma \tau/2})}\nonumber\\
\moy{\Pi_{I,1}}_w(t) &=& \frac{e^{-\Gamma (\tau-t)/2}-e^{-\Gamma \tau/2} }{2(1-e^{-\Gamma \tau/2})}\nonumber\\
\moy{\Pi_{II,0}}_w(t) &=& \frac{1}{1-e^{-\Gamma\tau}}.\nonumber\\
\moy{\Pi_{I,1}}_w(t) &=& 0.\nonumber\\
\ee
One can then deduce the weak values of the rank 2 projectors $\moy{\Pi_{I}}_w(t) = \moy{\Pi_{I,0}}_w(t)+\moy{\Pi_{I,1}}_w(t) =\moy{\Pi_{I,\mathrm{in}}}_w(t)+\moy{\Pi_{I,\mathrm{out}}}_w(t) =-1/(1-e^{\Gamma\tau})\simeq 0$ for $\Gamma\tau\gg 1$, $\moy{\Pi_{II}}_w(t)=\moy{\Pi_{II,0}}_w(t)+\moy{\Pi_{II,1}}_w(t) = 1/(1-e^{-\Gamma\tau}) \simeq 1$.

We then compute the energetic weak-values. Using $H_\mathrm{m}=\hbar\omega_\mathrm{m}\ket{1}_\mathrm{m}\bra{1}$, one has:
\bb
\moy{H_\mathrm{m}}_w(t) &=& \moy{H_\mathrm{m}\Pi_I}(t)\nonumber\\
 &=& \hbar\omega_\mathrm{m}\frac{e^{-\Gamma (\tau-t)/2}-e^{-\Gamma \tau/2} }{2(1-e^{-\Gamma \tau/2})}\\
\moy{H_\mathrm{m}\Pi_{II}}(t) &=&  0.
\ee
Similarly, one has:
\bb
\moy{H_\mathrm{ph}\Pi_I}_w(t) &=& \hbar\omega_\mathrm{ph}\moy{\Pi_{I,0}}_w(t)\nonumber\\&&+\hbar(\omega_\mathrm{ph}-\omega_\mathrm{m})\moy{\Pi_{I,1}}_w(t)\\
\moy{H_\mathrm{ph}\Pi_{II}}_w(t) &=& \hbar\omega_\mathrm{ph}\moy{\Pi_{II,0}}_w(t).
\ee

\section*{Appendic D: Backward propagation} 

In Ref.\cite{Vaidman2003Mar}, it is pointed out that the interaction-free nature of the measurement process is characterized by a vanishing wavefunction of the photon in the arm $I$ when propagated backward from the dark port. Within our framework, one can compute the evolution of the postselected joint bomb photon state $\ket{\tilde\Psi_\mathrm{f}} = \ket{\mathrm{dk}}_\mathrm{ph}\ket{\mathrm{in}}_\mathrm{m}\ket{0}_\mathrm{b}$. Using the same method as in the second section of the Appendix, we define the ansatz wavefunction:
\bb
\ket{\tilde\Psi_\mathrm{bw}(t)} &=&  \sum_j\bigg[ \Big(c'_0(\tau)\phi_0(\omega_j)+d'_0(\tau)\phi_{0}(\omega_j-\omega_\mathrm{m})\Big)\ket{0}_\mathrm{m}\nonumber\\
&&\!\!\!\!\!\!\!\!\!\!\!\!\!\!\!\!\!\!\!\! + \Big(c'_1(\tau)\phi_0(\omega_j+\omega_\mathrm{m})+d'_1(\tau)\phi_0(\omega_j)\Big)\ket{1}_\mathrm{m}\bigg]a_{I,j}^\dagger\ket{0}_\mathrm{ph}\ket{0}_\mathrm{b}\nonumber\\
&&\!\!\!\!\!\!\!\!\!\!\!\!\!\!\!\!\!\!\!\!-\frac{1}{\sqrt 2}\ket{II}_\mathrm{ph}\ket{\mathrm{in}}_\mathrm{m}\ket{0}_\mathrm{b},\quad
\ee
and look for the evolution equation for the coefficients $c'_{0,1}(\tau)$, $d'_{0,1}(\tau)$. We obtain:
\bb
\dot{{c}'}_0 &=& -\frac{\Gamma}{4}(\tilde c_0-\tilde c_1)\\
\dot{{c}'}_1 &=& \frac{\Gamma}{4}(\tilde c_0-\tilde c_1)\\
\dot{{d}'}_0 &=& -\frac{\Gamma}{4}(\tilde d_0-\tilde d_1)\\
\dot{{d}'}_1 &=& \frac{\Gamma}{4}(\tilde d_0-\tilde d_1).
\ee
Using the initial condition $({c'}_0(0),{c'}_1(0),d'_0(0),d'_1(0)) = (1/2,0,0,-1/2)$ associated with $\ket{\tilde\Psi_\mathrm{f}}$, we obtain for $\Gamma\tau\gg 1$:
\bb
c'_{0,1}(\tau)= -c'_{0,1}(\tau) \simeq \frac{1}{2\sqrt 2},
\ee
corresponding to state:
\bb
\ket{\tilde\Psi_\mathrm{bw}(\tau)} &\simeq& \frac{1}{\sqrt 2}\bigg(\sum_j\bigg[ \frac{ \phi_0(\omega_j)-\phi_{0}(\omega_j-\omega_\mathrm{m})}{2}\ket{0}_\mathrm{m}\nonumber\\
&& + \frac{ \phi_0(\omega_j+\omega_\mathrm{m})-\phi_0(\omega_j)}{2}\ket{1}_\mathrm{m}\bigg]a_{I,j}^\dagger\ket{0}_\mathrm{ph}\nonumber\\
&&-\ket{II}_\mathrm{ph}\ket{\mathrm{in}}_\mathrm{m}\bigg)\ket{0}_\mathrm{b}.
\ee

Now, the amplitude present in the arm $I$ is non zero due to the discrepancy between wavefunctions $\phi_{0}(\omega_j\pm\omega_\mathrm{m})$ and $\phi_{0}(\omega_j)$, which vanishes in the limit of large initian photon frequency variance $\Delta\omega\gg \omega_\mathrm{m}$. Specifically we have:
\bb
\vert\phi_0(\omega_j)-\phi_{0}(\omega_j-\omega_\mathrm{m})\vert &\simeq& \vert \phi_0(\omega_j)-\phi_{0}(\omega_j+\omega_\mathrm{m})\vert \nonumber\\
& \lesssim & \frac{\omega_\mathrm{m}}{\Delta \omega}.
\ee

\end{document}